\documentclass[10pt,letterpaper]{article}
\usepackage{opex3}
\usepackage{color}
\usepackage{threeparttable}

\usepackage{graphicx}
\usepackage{epsfig}
\usepackage{color}
\usepackage[hypertex,unicode=true]{hyperref}

\hypersetup{
        colorlinks=false,
        citecolor=blue,
        linkcolor=black,
        urlcolor=blue}

\newcommand{\halfl}{\ensuremath{{\scriptstyle \frac{1}{2}}}}

\newcommand{\un}   [1]{\ensuremath{\,\mathrm{#1}}}
\newcommand{\unit} [1]{\un{#1}}

\newcommand{\pder} [2]{\ensuremath{\frac{\partial #1}{\partial #2}}}
\newcommand{\pderl}[2]{\ensuremath{\partial #1/\partial #2}}

\newcommand{\ket}  [1]{\ensuremath{\left | #1 \right \rangle }}

\newcommand{\neff} {\ensuremath {n_{\mathrm{eff}}}}
\newcommand{\Lint} {\ensuremath {L_{\mathrm{int}}}}
\newcommand{\sint} {\ensuremath {s_{\mathrm{int}}}}
\newcommand{\Pref} {\ensuremath {P_{\mathrm{ref}}}}
\newcommand{\Pin}  {\ensuremath {P_{\mathrm{in}}}}
\newcommand{\Ct}   {\ensuremath {C_{\mathrm{target}}}}

\begin{document}

\title{Design and characterization of integrated components for SiN photonic quantum circuits}

\author{Menno~Poot, Carsten~Schuck, Xiao-song~Ma, Xiang~Guo, and Hong~X.~Tang$^*$}
\email{*hong.tang@yale.edu}\address{Department of Electrical
Engineering, Yale University, New Haven, CT 06520, USA}


\begin{abstract*}
The design, fabrication, and detailed calibration of essential building blocks towards fully integrated linear-optics quantum computation are discussed. Photonic devices are made from silicon nitride rib waveguides, where measurements on ring resonators show small propagation losses. Directional couplers are designed to be insensitive to fabrication variations. Their offset and coupling lengths are measured, as well as the phase difference between the transmitted and reflected light. With careful calibrations, the insertion loss of the directional couplers is found to be small. Finally, an integrated controlled-NOT circuit is characterized by measuring the transmission through different combinations of inputs and outputs. The gate fidelity for the CNOT operation with this circuit is estimated to be 99.81\%  after post selection. This high fidelity is due to our robust design, good fabrication reproducibility, and extensive characterizations.
\end{abstract*}



\section{Introduction}
Single photons combined with linear optical elements and efficient detectors show tremendous promise for universal quantum computation \cite{knill_nature_quantum_computation_linear_optics, kok_RMP_LOQC, nielsen_book_quantum_computation}. In linear-optics quantum computation (LOQC) quantum information is carried by single photons which are manipulated using linear optical elements such as beam splitters, waveplates, and phase shifters. The first steps towards the goal of universal quantum computation \cite{kok_RMP_LOQC, nielsen_book_quantum_computation} have been achieved, see e.g. \cite{obrien_PRL_CNOT_tomography, walther_nature_oneway, pittman_PRA_error_correction, okamoto_PNAS_CNOT_KLM, gao_PNAS_teleportation_gate}. However, scaling these experiments to larger system sizes is difficult in bulk optics due to its limited mechanical stability and space requirements. On the other hand, integrated photonic circuits easily achieve scalability, compactness, and interferometric stability \cite{politi_science_quantum_circuits}. Here we discuss the design, fabrication, and characterization of photonic circuits for on-chip LOQC based on an integrated silicon nitride platform. Our vision is to integrate all the required components for LOQC on a single chip to minimize coupling (i.e. interconnect) losses. This includes sources of single photons \cite{lounis_RPR_single_photon,eisaman_RSI_SSPD_single_photons}, programmable circuitry for realizing quantum-logic operations \cite{broome_science_boson_sampling, carolan_science_universal_LOQC}, and detection of the single photons that make up the qubits \cite{eisaman_RSI_SSPD_single_photons, natarajan_SST_SSPD_review, pernice_natcomm_SSPD}.

There are different ways to encode quantum information using photons, for example using their polarization, or through time-bin encoding \cite{nielsen_book_quantum_computation}. However, in integrated photonic circuits, a natural way is to encode quantum information in the spatial mode of the single photon \cite{kok_RMP_LOQC}. In this framework the logical states of the qubit (written as $\ket{1}$ and $\ket{0}$) correspond to which (single-mode) waveguide a single photon travels through. Single-qubit operations then are implemented using relative phase shifts and directional couplers. Another essential element for quantum computation are multi-qubit operations such as controlled-NOT (CNOT), controlled-Z, and Toffoli gates \cite{nielsen_book_quantum_computation}.
Finally, after (or even during) the computation the single photons have to be detected with high efficiency. For this purpose superconducting nanowire single photon detectors (SNSPDs) are an excellent choice because of their good quantum efficiency at near-infrared wavelengths, and high counting speed \cite{natarajan_SST_SSPD_review}.

In our vision, all the required components will be fabricated on a single monolithically-integrated chip. This thus also includes the SNSPDs, which can be fabricated directly on top of the photonic waveguides \cite{pernice_natcomm_SSPD, schuck_APL_OTDR,sprengers_APL_waveguide_SSPD,
schuck_SR_NbTiN_SSPD_dark_count}. In this way, coupling losses between the quantum circuitry and the detectors that would occur if SNSPDs were situated on a separate chip \cite{natarajan_SST_SSPD_review}, are avoided. However, since such completely integrated chips have to be measured at cryogenic temperatures, the on-chip power dissipation should be small. For this purpose, we have developed opto-electromechanical phase shifters \cite{poot_apl_phaseshifter,poot_H_optomechanics}, which are also used for optomechanics experiments \cite{poot_PRA_squeezing_feedback, poot_NJP_Yfeedback, rath_APL_diamond_Hresonator,st-gelais_NL_Hresonator_heat_transfer}.
Currently, the nonclassical light for our experiments is generated off the chip using spontaneous parametric down-conversion (SPDC) of 777 nm light in a periodically-poled potassium-titanyl-phosphate (pp-KTP) waveguide \cite{schuck_natcomm_HOM}. Future integration of the source can be done, for example, by using deposited AlN \cite{xiong_NJP_AlN_integrated_optics}, bonded GaN \cite{xiong_OE_GaN_SHG}, or LiNbO$_3$ \cite{wang_OE_LiNbO3_disks} as the source material for SPDC. Alternatively, photon pairs generated through spontaneous four-wave mixing can be used, e.g. \cite{sharping_OE_Si_FWM_pairs, takesue_APL_FWM_pairs}
Unlike SPDC this process is also possible in materials with inversion symmetry, including SiN \cite{dutt_PRAp_onchip_squeezing}.

After our previous demonstrations of phase shifters and single photon detectors, in this work we focus on the photonic circuits themselves. Section \ref{sec:fabrication} discusses the choice for the SiN material platform and the fabrication of the circuits. It is shown in Sec. \ref{sec:propagationloss} that the resulting waveguides have low propagation loss. The directional couplers are the subject of Sec. \ref{sec:photonic} to \ref{sec:phase_shift}: Their photonic design is discussed in Sec. \ref{sec:photonic} and the characterization in Sec. \ref{sec:directional}. The interaction loss is the topic of Sec. \ref{sec:insertionloss} and the phase between the outputs of the directional couplers is studied in Sec. \ref{sec:phase_shift}. Finally, in Sec. \ref{sec:CNOT_ports} a complete quantum-operation circuit is measured and characterized using the transmission of classical light.

\section{Photonic design and fabrication}
\label{sec:fabrication}
In our integration approach, our photonic quantum-optics platform includes mechanically active elements, nanophotonic waveguides, as well as thin superconducting detectors. The choice of materials should thus be compatible with the release of the mechanical parts, allow for low-loss waveguides (as the loss of a photon would destroy the qubit), and enable deposition of high-quality superconducting material. Taking these considerations into account leads to the conclusion that silicon nitride is the best candidate for our platform. High-stress SiN has excellent mechanical properties as well as good optical transmission for near-infrared light \cite{gondarenko_OL_highQ_SiN_ring, tien_OL_highQ_SiN_ring}. Moreover, it is compatible with the sputter-deposition of NbTiN for highly efficient SNSPDs \cite{schuck_SR_NbTiN_SSPD_dark_count}.

After selecting SiN as the material of choice, a fabrication process has to be devised which should be accurate and reproducible, yield low-loss waveguides, and be compatible with releasing the mechanical structures, without degrading the superconducting material. The devices in this work only contain the optical circuitry and can thus be made in a simpler, but fully compatible, process flow. However, since the detectors and phase shifters are essential elements in our fully integrated programmable photonic platform, we will now first outline the entire fabrication process envisioned for photonic chips with opto-electromechanical phase shifters and SNSPDs (``full devices'') before focusing on the simplified process that is used for the devices presented here. For full devices, the starting point will be a silicon wafer with an oxide cladding layer and a SiN device layer in which the photonic circuits will be defined. After deposition of the superconducting material, the entire fabrication scheme for a full device including electromechanical phase shifters and superconducting detectors will consist of 6 lithography steps. First, metallic markers and electrodes will be defined by electron-beam lithography and evaporation. Then the SNSPDs will be written and etched. Next a protective layer will be placed over the alignment markers and then the silicon nitride is to be etched in two separate steps: one for the mechanical structures that have to be released, and another one where a thin layer of SiN remains. This layer will act as an etch mask that protects the underlying silicon oxide during the release of mechanical structures \cite{poot_apl_phaseshifter, poot_H_optomechanics}. In the final step, etch windows around the optomechanical phase shifters will be opened in photoresist such that only the mechanical structures will be exposed to buffered hydrofluoric acid for release. The remaining photoresist will protect the NbTiN detectors during the release and is to be removed afterwards. This is the process flow that we envision for the future generation of completely integrated quantum-optics chips.

On the other hand, without movable parts and SNSPDs, the photonic devices presented in this work are made in a single lithography step. A silicon chip with a $3.3 \un{\mu m}$ thick silicon oxide cladding and a 330 nm thick high-stress SiN device layer is cleaned and spin-coated with ZEP520A resist. The patterns are defined by electron beam lithography using a Raith EBPG5000+. After development in xylene, the nanophotonic structures are etched using a CHF$_3$/O$_2$ plasma. The etch is timed such that $\sim 50 \un{nm}$ of SiN remains; the photonic waveguides thus have a so-called rib structure.

To couple light onto and out of the chip, grating couplers are etched into the SiN \cite{taillaert_IEEE_grating_couplers}. By placing an array of single-mode optical fibers above such structures, light is coupled from the fibers into the waveguides and vice versa. The waveguides support a single TE-like mode and the polarization of the input light is adjusted using a fiber polarization controller. Since the sample can move freely underneath the fiber array (that holds eight fibers), it is easy to sequentially measure the large number of devices required for this work.

\section{Propagation loss measurements}
\label{sec:propagationloss}
\begin{figure}[tb]
\includegraphics{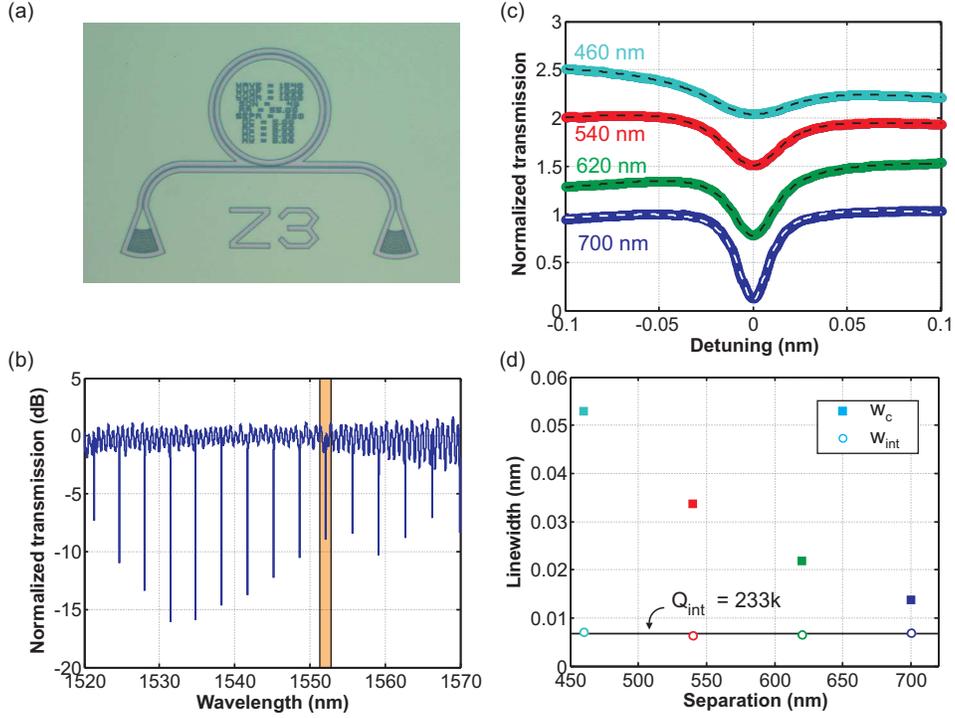}%
\caption{(a) Optical micrograph of a ring resonator with a feeding waveguide and grating couplers (triangular structures). The spacing between the grating couplers is 250 um and the ring has a diameter of 110 um.
(b) Optical transmission of a device with a waveguide width of 1 um and a separation between the ring and the feeding waveguide of 700 nm. The transmission is normalized to that of the grating couplers. The shaded area indicates the resonance highlighted in panel (c). 
(c) Zooms of resonances near 1552 nm with fits (dashed lines) for different separations between the waveguide and the ring on a linear scale. The curves are offset for clarity.
(d) Linewidths extracted from the fits. The solid symbols are the linewidths due to the coupling between the feed waveguide and the ring, whereas the open circles are the internal linewidths. The colors are consistent between all panels.
\label{fig:rings}
}
\end{figure}

As explained above, the photonic circuits for LOQC need to have low optical losses as absorption or scattering of a photon would destroy the qubit. Rib waveguides made out of SiN can have low propagation losses \cite{melchiorri_APL_SiN_propagation_loss, gruhler_OE_SiN_graphene}, which is also true for our devices as we will show next. Figure \ref{fig:rings}(a) shows a picture of a test device to quantify the propagation loss. It consists of a ring resonator that is coupled to a waveguide connecting two grating couplers for optical input and output. The resonances of the ring appear in the optical transmission spectrum, as shown in Fig. \ref{fig:rings}(b). Their spacing is used to extract the group index $n_g$, whereas their linewidths determine the propagation loss. A free-spectral range of FSR = 3.465 nm is found by fitting the resonances. From this value the group index can be obtained using $n_g = \lambda^2/(2\pi R \times \mathrm{FSR})$ \cite{bogaerts_LPR_ring_review}. Here $R = 55 \un{\mu m}$ is the ring radius and $\lambda = 1550 \un{nm}$ is the vacuum wavelength. This is done for four different devices with different separations between the waveguide and the ring. The group index is independent of this distance and has a mean value $n_g = 1.9936 \pm 0.0017$. This value is typical for such waveguides \cite{gruhler_OE_SiN_graphene}. On the other hand, the resonances themselves do depend on the ring-waveguide separation. This is because the coupling between the waveguide and the ring, as quantified by the coupling linewidth $w_c$, increases with decreasing separation. The total width of the resonance is the sum of $w_c$ and the linewidth due to internal losses such as scattering and absorption, $w_{\mathrm{int}}$.
Note that $w_{\mathrm{int}}$ does not depend on the separation, so that by decreasing the separation the resonances go from undercoupled ($w_c \ll w_{\mathrm{int}}$) through critically coupled ($w_c = w_{\mathrm{int}}$) to overcoupled ($w_c \gg w_{\mathrm{int}}$). The extinction, i.e. the depth of the resonances, has a maximum at critical coupling. As shown in Fig. \ref{fig:rings}(c) for small separation (cyan) the dip is shallow and wide, which is characteristic for an overcoupled resonance. For larger separations the dip is much narrower and has a higher extinction. From input-output theory the shape of the resonance is expected to be Lorentzian \cite{bogaerts_LPR_ring_review}. However, Figure \ref{fig:rings}(b) shows that the (off-resonant) transmission is not flat: it contains Fabry-P\'erot fringes due to reflection at the grating couplers. The fit function is the product of these two contributions:
\begin{equation}
T(\lambda) = \frac{T_0}{1+F_c \sin^2(\pi[\lambda-\lambda_{\mathrm{fp}}]/\mathrm{FSR}_{\mathrm{fp}})} \times \left(1-\frac{w_c w_{\mathrm{int}}} {(w_c+w_{\mathrm{int}})^2/4+(\lambda - \lambda_0)^2}\right),\label{eq:fit_io}
\end{equation}
where $T_0$, $F_c$, $\lambda_{\mathrm{fp}}$, and $\mathrm{FSR}_{\mathrm{fp}}$ are the fit parameters for the fringes, and $\lambda_0$, $w_{\mathrm{int}}$, and $w_c$ are those for the actual resonance.
The dashed lines in Fig. \ref{fig:rings}(c) are the resulting fits; they describe the data well and give accurate values for the resonance wavelength $\lambda_0$ and both the internal and couplings linewidths $w_{\mathrm{int}}$ and $w_c$, respectively. However, the same fit is obtained with their values interchanged, as Eq. (\ref{eq:fit_io}) is symmetric with respect to these two parameters. Hence, for a single ring it is not possible to distinguish between these two contributions (unless one has access to the phase of the transmitted light), but they can be identified via their dependence on the separation between the ring and the feeding waveguide [Fig. \ref{fig:rings}(d)]. The larger of the two linewidths (solid markers) clearly decreases with increasing separation, whereas the smaller of the two (open symbols) is independent of the gap. The former is thus identified as the coupling linewidth, whereas the latter is due to the internal losses. Its mean value of $w_{\mathrm{int}} = 6.65 \unit{pm}$ yields an intrinsic quality factor $Q_{\mathrm{int}} = \lambda_0/w_{\mathrm{int}} = 2.33\times10^5$. As mentioned above, $Q_{\mathrm{int}}$ is directly related to the propagation loss of the ring by $Q_{\mathrm{int}} = 2\pi n_g/\alpha\lambda$ \cite{rabiei_JLT_polymer_rings, rath_OE_diamond_nanophotonics_waferscale}, where $\alpha$ is the propagation loss coefficient and $\lambda$ the free-space wavelength. Inserting the values for $Q_{\mathrm{int}}$ and $n_g$ gives $\alpha = 0.35 \un{cm^{-1}}$ which equals $1.5 \un{dB/cm}$. This shows that for the devices shown in this work, where the light travels through less than 1 mm of waveguide, propagation losses are negligible. However, we note the loss can be reduced even further by reflowing the developed resist before etching the photonic structures \cite{fong_NL_water_wheel, gruhler_OE_SiN_graphene}. This smoothes the waveguide, thus reducing the scattering loss even further. After characterizing the loss in isolated waveguides, next the design of the directional coupler, another important element for LOQC, is discussed.

\section{Directional coupler design}
\label{sec:photonic}
\begin{figure}[tb]
\includegraphics{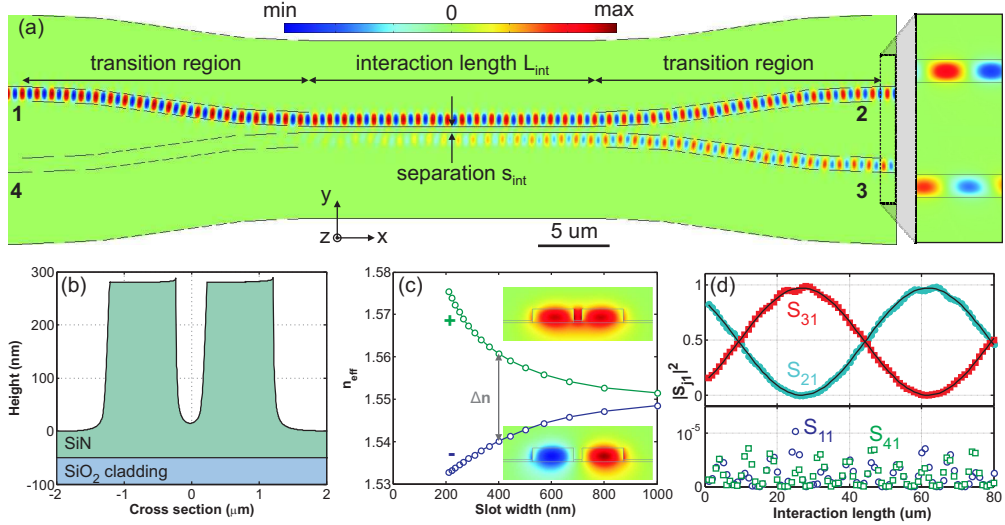}%
\caption{(a) Finite element simulation of a directional coupler (top view) when light is sent into port 1. The z-component of the magnetic field $H_z$ is shown according to the color scale. The inset shows a zoom of $H_z$ near the outputs (ports 2,3).
(b) Measured atomic-force microscope profile (averaged) across the center of a directional coupler with 1000 nm wide waveguides and 400 nm separation between them. The image is colored to indicate the guiding and cladding layer. The distance between the top of the waveguide and the cladding is 330 nm. Note the difference in the horizontal and vertical scales.
(c) Calculated effective refractive index of the even (green) mode and the odd (blue) mode versus the separation between the two waveguides. The insets show the $E_y$ field of the modes.
(d) Magnitudes of the elements of the scattering matrix for different $\Lint$ obtained from FEM simulations as in panel (a).
\label{fig:photonic}}
\end{figure}

Beam splitters are very important elements for linear quantum optics circuits when using free space optics. Together with phase shifters they can be used to generate any unitary operation \cite{reck_PRL_unitary}. In integrated photonics the device equivalent to a beam splitter is called a directional coupler. As shown in Fig. \ref{fig:photonic}(a) it consists of two waveguides that are initially well separated, but over a transition region they are brought in close proximity to each other. In this central area with interaction length $\Lint$ the waveguides run parallel to each other with separation $\sint$, and they are close enough so that the evanescent fields couple their optical modes \cite{huang_JOSAA_coupled-mode_theory}. Figure \ref{fig:photonic}(b) shows a cross-section of an actual directional coupler measured with an atomic force microscope. By simulating this structure through finite-element (FEM) simulations, the two guided optical modes are found as shown in Fig. \ref{fig:photonic}(c). For large slot widths their effective indices are similar and approach the effective refractive index of an isolated waveguide: $\neff \rightarrow 1.548$.
However, when the slot is narrowed, the mode coupling increases and the effective indices split into an upward ($n_{\mathrm{eff},+}$) and downward ($n_{\mathrm{eff},-}$) branch. The insets show that these modes correspond to the symmetric (even) and anti-symmetric (odd) superpositions of the waveguide modes, respectively. When light is sent into one port, say the upper right arm of the directional coupler in Fig. \ref{fig:photonic}(a), the light is initially localized in the mode of that waveguide, which is an equal superposition of the even and odd eigenmodes. The light in those modes then propagates along the length of the directional coupler. However, due to the difference in effective refractive indices $\Delta n = n_{\mathrm{eff},+} - n_{\mathrm{eff},-}$, a phase difference develops between the light in the even and odd mode. This phase difference is important when projecting the eigenmodes back onto the uncoupled waveguide modes for analysis at the outputs of the directional coupler. A full power transfer from one waveguide to the other results when the acquired phase difference is $\pi$. By varying the interaction length $\Lint$ of the directional coupler, the phase difference, and hence the power splitting ratio, can be controlled.

The analysis of the directional coupler using the effective refractive indices does not take the transition regions into account. To include these, finite-element simulation as shown in Fig. \ref{fig:photonic}(a) are performed. The amplitude and phase of the output fields, and thus the power-splitting ratio, are related to the input fields by the scattering matrix $\mathbf{S}$. The device has four ports and thus $\mathbf{S}$ is in principle a $4 \times 4$ matrix. However, similarly to a bulk optic beam splitter, hardly any light is reflected backwards [that is, back to port 1 or 4 for light inserted into port 1 as in Fig. \ref{fig:photonic}(a)]. This is confirmed by the simulation results shown in the lower panel of Fig. \ref{fig:photonic}(d), where the magnitudes  $|S_{11}|^2$ and $|S_{41}|^2$ are found to be below $10^{-5}$. It is then often convenient to identify the input and output ports of the directional coupler and use a reduced $2 \times 2$ scattering matrix  $\mathbf{S'}$ to relate those fields:
\begin{equation}
\left(\begin{array}{c}
a_2\\a_3
\end{array}\right)_{\mathrm{out}} = \left(\begin{array}{cc}
S_{21} & S_{24} \\ S_{31} & S_{34}
\end{array}\right)
\left(\begin{array}{c}
a_1\\a_4
\end{array}\right)_{\mathrm{in}}
\leftrightarrow
\left(\begin{array}{c}
a_1\\a_2
\end{array}\right)'_{\mathrm{out}} = \left(\begin{array}{cc}
S'_{11} & S'_{12} \\ S'_{21} & S'_{22}
\end{array}\right)
\left(\begin{array}{c}
a_1\\a_2
\end{array}\right)'_{\mathrm{in}}.
\label{eq:Smatrix}
\end{equation}
Here $|S'_{11}|^2$ indicates how much power of the input light (with power $\Pin$) simply goes through the original waveguide ($P_t$) and $|S'_{21}|^2$ indicates how much crosses over ($P_c$). From coupled-mode theory \cite{huang_JOSAA_coupled-mode_theory} it is found that
\begin{equation}
\frac{P_c}{\Pin} \equiv C = |S'_{21}|^2 = \sin^2\left(\frac{\pi}{2} \frac{\Lint+\ell_0}{\ell_c} \right).\label{eq:directional}
\end{equation}
Here, $\ell_0$ is the offset length which takes into account the power transferred in the transition region. Moreover, $\ell_c = \halfl \lambda/\Delta n$ is the coupling length. Note that both $\ell_c$ and $\ell_0$ do not correspond to any physical dimension of the coupler. They both depend on $\sint$ and on the wavelength $\lambda$: a smaller $\sint$ results in a stronger coupling between the waveguide modes and thus in a shorter $\ell_c$. To design a directional coupler with a certain target splitting ratio $\Ct$, one can either select $\sint$, which fixes $\ell_c$ and $\ell_0$, and then find the required $\Lint$ using Eq. (\ref{eq:directional}). Alternatively, one can also first fix a value for $\Lint$ and then (numerically) solve for $\sint$, but, as discussed below, the former approach is used for designing our devices. Graphically, the solutions are the intersections of the red curve in Fig. \ref{fig:photonic}(d) with a horizontal line at $C = \Ct$. For each maximum of the sinusoidal curve there are two intersections with different signs of their slopes. The maxima themselves can be labeled by an integer $k = \mathrm{round}(\{\Lint + \ell_0\}/2\ell_c)$ that counts the number of times light has been transferred back and forth between the two waveguides after traversing the directional coupler. Inverting Eq. (\ref{eq:directional}) yields the solutions $\Lint = \ell_c [2k \pm \arccos(1-2 C)/\pi] - \ell_0$. It can be seen in Fig. \ref{fig:photonic}(d) that solutions labeled with a + (-) sign have a negative (positive) slope, and that solutions with higher $k$ correspond to larger device lengths. By taking the derivative of Eq. (\ref{eq:directional}) with respect to $\lambda$, the wavelength dependence of the splitting ratio is found:
\begin{equation}
\left|\pder{C}{\lambda}\right| = \frac{1}{\ell_c}\sqrt{C(1-C)} \left\{\pi \pder{\ell_0}{\lambda} - \left[2\pi k \pm \arccos(1-2C)\right]\pder{\ell_c}{\lambda} \right\}.\label{eq:direc_dispersion}
\end{equation}
From FEM simulations it is found that $\pderl{\ell_0}{\lambda}$ is positive and an order of magnitude smaller than the negative-valued $\pderl{\ell_c}{\lambda}$. This means that the lowest value of $k$ for which there is a positive $\Lint$ results in the smallest wavelength dependence, i.e. when working on the first period of the oscillations of the power between the two waveguides [cf. Fig. \ref{fig:photonic}(d)].

\begin{table}[tbp]
\centering \caption{
Changes of the coupling length $\delta\ell_c$ from the nominal value $\ell_c = 37.47 \un{\mu m}$ for variations of the device parameters. The values of the variations (third column) represent estimated fabrication uncertainties around the target values (second column). The ``remaining thickness'' is the thickness of the SiN remaining on top of the SiO$_2$ after etching, far outside the coupler region. The ``center thickness'' is the thickness remaining in between the two waveguides, which is larger due to the narrow slot. All variations $\delta p$ are small enough for the changes in coupling length to be proportional to the variation in the parameter $p$ that caused it: $\delta \ell_c \approx \pderl{\ell_c}{p} \times \delta p$.} \label{tab:sensitivities}
\hspace{-2cm}
\begin{tabular}{|cccr|}
\hline
{\bf Quantity} & {\bf Value} & {\bf Variation} &
$\mathbf{\delta \ell_c  (\un{\mu m})}$ \\
\hline
width left waveguide  & 1000 nm & -25 nm 
& -1.15 \\
width right waveguide & 1000 nm & -25 nm 
& -1.15 \\
slot width $\sint$    &  400 nm & +25 nm 
&3.29 \\
lateral etch & - & +25 nm 
& 1.85 \\
SiN thickness & 330 nm & -10 nm 
&-1.46 \\
remaining thickness & 50 nm & -10 nm 
& -0.62 \\
center thickness & 70 nm & -10 nm 
& 2.06 \\
ref. index SiN & 2.00 & 0.01 
&1.07 \\
ref. index SiO$_2$ & 1.44 & 0.01 
&-0.33 \\
sidewall angle & $17\deg$ & $1\deg$ 
&-0.26 \\
wavelength $\lambda$ & 1550 nm & 5 nm 
&-0.59 \\
\hline
\end{tabular}
\end{table}

As explained above, using either FEM simulations or measurements on previous devices, the design parameters for a directional coupler with a certain splitting ratio can be calculated. In practice, it is hard to exactly match the target dimensions for the directional coupler and also variations between fabrication runs occur. However, as will be shown in Sec. \ref{sec:CNOT_ports} the splitting ratios $C$ have to be close to their target values for performing high-fidelity operations. Therefore, it is important to see how the coupling length $\ell_c$ (which is the most important contribution; see the discussion on the wavelength dependence above) changes due to differences in the coupler geometry. The smaller these changes $\delta \ell_c$ are, the more reproducible the directional couplers will be. To optimize these, first a FEM simulation is performed with device parameters (second column of Table \ref{tab:sensitivities}) that reflect the cross section of fabricated directional couplers [Fig. \ref{fig:photonic}(b)]. For this nominal geometry the resulting coupling length is $\ell_c = 37.47 \un{\mu m}$. Now a change in $\ell_c$ can be converted to a change in $C$ using Eq. (\ref{eq:directional}). For example, for a 50/50 directional coupler it is found that $C$ changes by -0.021 per micrometer change in $\ell_c$.

After simulating the nominal device, simulations are performed with small changes in the device geometry. Table \ref{tab:sensitivities} shows the resulting changes $\delta \ell_c$. As an example, consider the effect of changes in the width of the left waveguide (first row). Its designed width is 1000 nm but if the waveguide turns out to be 25 nm narrower, the coupling length decreases by $1.15 \un{\mu m}$. This is because the modes will be less confined in a narrower waveguide, which increases the coupling between the modes. Hence, $\Delta n$ increases from 0.02069 to 0.02135, resulting in the shorter $\ell_c$. The effects of other variations are given in the other rows of Table \ref{tab:sensitivities}. It is clear that $\ell_c$ is most sensitive to the slot width. In practice the waveguide and slot widths, as defined in electron beam lithography, are very accurate; variations mainly arise during the dry etch of SiN. A small amount of lateral etching reduces the widths of both waveguides and at the same time increases the width of the slot. In our device geometry these contributions have opposite effects on $\ell_c$ so that they largely cancel each other: 25 nm of lateral etching on both sides of each waveguide only increases the coupling length by $1.85 \un{\mu m}$. This number is to be compared to the situation where both waveguides are 50 nm narrower but where the slot width stays the same; in that case $\ell_c$ changes much more: $\delta \ell_c = 4\times-1.15 \un{\mu m} = -4.6 \un{\mu m}$. Similarly, when $\sint$ is 50 nm larger, $\delta \ell_c =  2\times 3.29 \un{\mu m} = +6.6 \un{\mu m}$. These opposite effects thus largely cancel each other for lateral etching, indicating that our design is quite robust against such variations. Another important variation can arise when the superconductor is removed in the fully integrated device (i.e. one with the superconducting detectors and phase shifters): Since the dry etch is not selective between NbTiN and SiN, a thin layer of the underlying SiN film will be removed in that process. As indicated in Table \ref{tab:sensitivities} a 10 nm thinner SiN film results in a $1.46 \un{\mu m}$ shorter $\ell_c$. However, if, despite the thinner film, the etch depth that defines the waveguides is kept the same, the thickness of the SiN remaining in the center of the slot will also be thinner, which increases $\ell_c$ by $2.06 \un{\mu m}$. Moreover, the remaining SiN away from the directional coupler is also thinned giving $\delta \ell_c = -0.62 \un{\mu m}$. Adding up these three contributions results in a vanishing change in $\ell_c$, again indicating the robustness of the chosen design.

\section{Characterization of the directional couplers}
\label{sec:directional}
Linear optics quantum circuits require directional couplers with well-defined splitting ratios. For example, a particular design of a CNOT gate (see Sec. \ref{sec:CNOT_ports}) requires directional couplers with $C = 1/2$ and $C = 2/3$ \cite{ralph_PRA_CNOT_coincidence_theory}. The closer the splitting ratios are to those ideal values, the higher the fidelity of the quantum operation is \cite{ralph_PRA_CNOT_coincidence_theory}. Getting as close as possible to the design values is thus extremely important. As discussed in the previous section, realizing a certain splitting ratio for a directional coupler means finding the right interaction length $\Lint$, which requires knowledge of $\ell_c$ and $\ell_0$. The simulations of Sec. \ref{sec:photonic} give a good indication of their expected values, but the agreement between simulation and experiment has to be checked in order to obtain the right splitting ratios that are required for high-fidelity operations.

\begin{figure}[tb]
\includegraphics{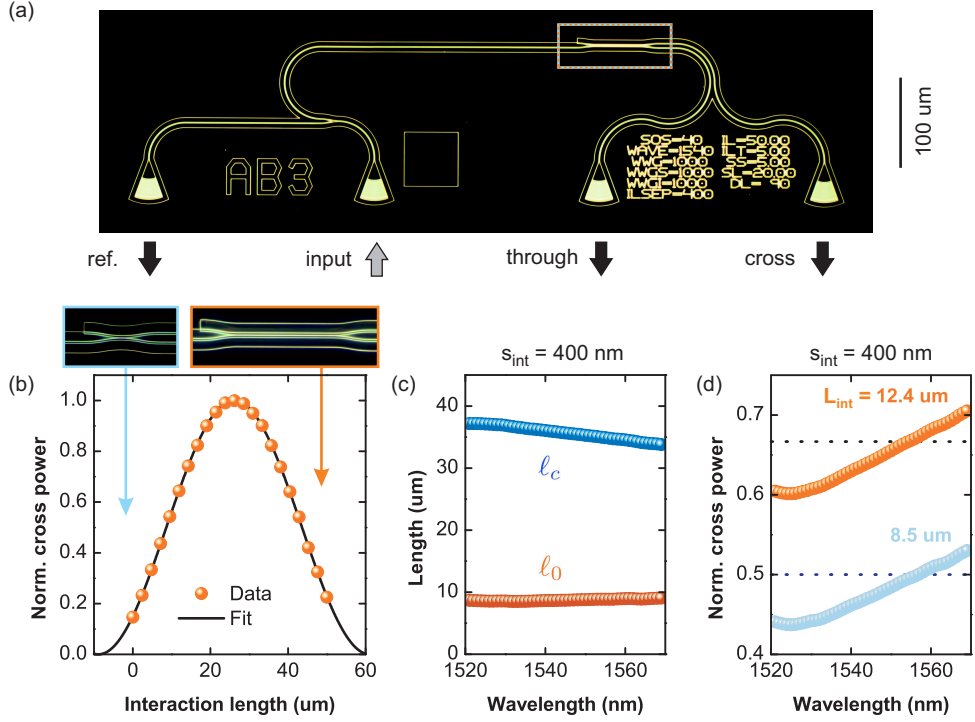}%
\caption{(a) Optical dark-field image of a device to calibrate
the coupling ratio of the directional coupler (dashed rectangle). The insets show two directional couplers with different interaction lengths.
(b) Measured normalized cross power $P_c/(P_t+P_c)$ at a wavelength of 1554 nm for different $\Lint$, together with a fit of Eq. (\ref{eq:directional}) (solid line) to the data.
The data is obtained by measuring the transmission profiles with a tunable laser, followed by a 2 nm averaging.
(c) The extracted wavelength dependence of the coupling length $\ell_c$ and the offset length $\ell_0$.
(d) The cross power as a function of wavelength for directional couplers with
two different $\Lint$ values. The dashed lines indicate 50/50 and 33/67 beam splitters.
%
}
\label{fig:directional_intlen}
\end{figure}

Using devices such as the one shown in Fig. \ref{fig:directional_intlen}(a) with varying $\Lint$, we experimentally obtain $\ell_c$ and $\ell_0$. Light is sent into the input (blue arrow) and is split equally using a Y-splitter \cite{zhang_OE_Ysplitter}. Half of the light goes to the reference output (left), whereas the other half is sent into the lower arm of the directional coupler where it is redistributed over the two outputs. The resulting powers $P_t$ and $P_c$ are then detected at the respective outputs. Figure \ref{fig:directional_intlen}(b) shows the normalized cross power $P_c/(P_c+P_t) = C$ for 22 different devices with $\Lint$ ranging from 0 to $50 \un{\mu m}$. For the shortest device only a small amount of power crosses over and the remaining light goes through the same waveguide. Initially, $C$ increases with increasing $\Lint$ until the maximum where all the light crosses over to the opposite waveguide is reached at $\Lint = 26 \un{\mu m}$. For even longer interaction lengths the cross power steadily drops, indicating that more light is transferred back to the original (``through'') waveguide. The solid line is a fit of Eq. (\ref{eq:directional}) and describes the data well. The period is $2\ell_c$, whereas $-\ell_0$ is the (negative) value where the function reaches zero. By repeating these fits at different wavelengths, the wavelength dependence of $\ell_c$ and $\ell_0$ shown in Fig. \ref{fig:directional_intlen}(c) is obtained. The offset length depends only weakly on $\lambda$, but the coupling length shows a clear decrease with increasing wavelength: at longer wavelengths the evanescent field extends further out from the waveguide. Hence the coupling is stronger, resulting in a shorter $\ell_c$. Figure \ref{fig:directional_intlen}(d) shows the wavelength-dependence of the cross power for two interaction lengths. At $\lambda = 1554 \un{nm}$, i.e., the wavelength where our SPDC source operates \cite{schuck_natcomm_HOM}, the coupling ratios are very close to the values of 1/2 and 2/3 that are required for the CNOT gate (dashed lines). When moving away from this center wavelength, deviations in the beam-splitting ratio become larger. However, within the bandwidth of the down-converted photons ($\sim 4.6 \un{nm}$  \cite{schuck_natcomm_HOM}), the splitting ratio only changes from $C_{1/2} = 0.489$ to $0.499$ and $C_{2/3} = 0.660$ to $0.671$ for $\Lint = 8.5$ and $12.4 \un{\mu m}$ respectively. The splitting ratios are thus less than 0.011 away from the ideal values over the entire wavelength range of interest. A further reduction of the dispersion could be achieved by adapting an asymmetric design for the directional coupler \cite{lu_OE_broadband_directional_coupler}.


Finally, we tested the reproducibility of our directional couplers by comparing chips that were fabricated months apart: The values of $\ell_0$ and $\ell_c$ for these runs differed by only $\sim 2 \%$, highlighting the robustness of our directional coupler design and fabrication.

\section{Insertion loss}
\label{sec:insertionloss}
Another important parameter of a directional coupler is its insertion loss, i.e., how many of the photons sent into the input do not come out of the two outputs. In other words: how many photons are lost, for example due to absorption, reflection, or scattering? To this end, the combined optical power at the output ports $P_t+P_c$ is compared with the power at the input of the directional coupler, which equals the power at the reference port $\Pref$.
\begin{figure}[tb]
\includegraphics{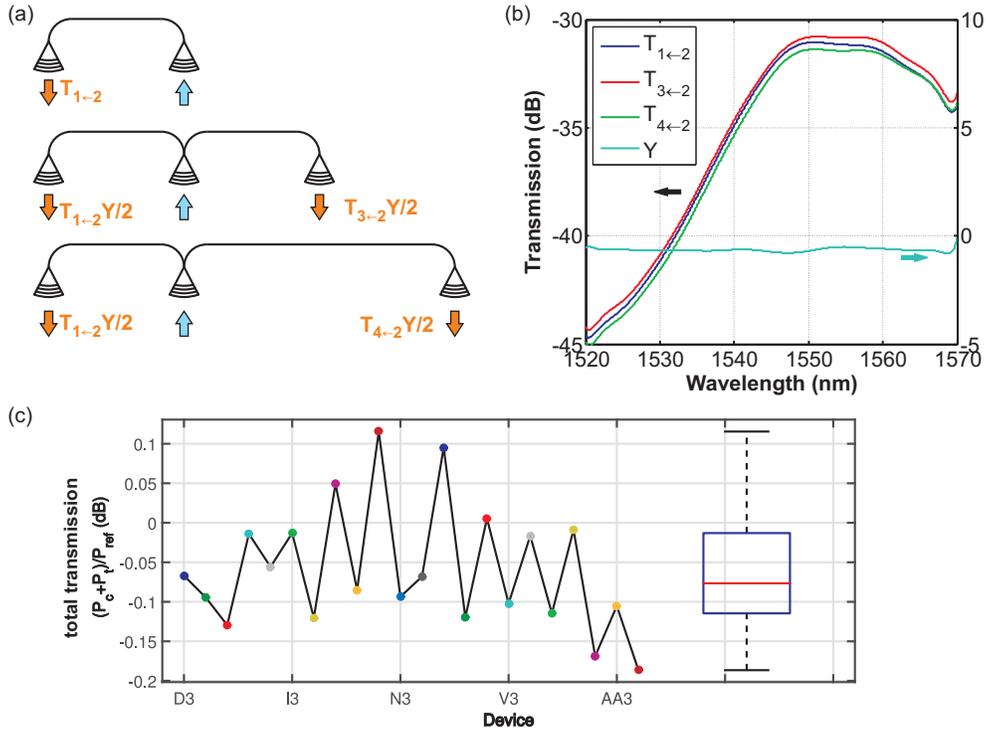}%
\caption{(a) Schematics of calibration devices to determine the insertion loss. Light is sent into port 2 (blue arrow) and detected at the other ports (orange). The three devices have different routings that enable extraction of the coupler transmissions $T_{i\leftarrow2}$ and the insertion loss of the Y-splitter $Y$. The combinations that the light encounters are indicated at each output grating coupler.
(b) Extracted transmission profiles $T_{i\leftarrow2}(\lambda)$ and $Y(\lambda)$.
(c) Insertion loss of the 22 directional couplers (same devices as Fig. \ref{fig:directional_intlen}, averaged over wavelength) and a box plot showing the median value (red), 25 and 75\% percentile (box) and the extent (bars). The interaction length is varied from 0 to 50 $\un{\mu m}$ when going from device D3 to AB3.
\label{fig:insertionloss}}
\end{figure}
Determining small insertion losses requires accurate measurements of the relative transmission through many devices on the chip. Hence such a measurement is only possible after careful calibration as there are several factors that need to be taken into account. First of all, the sample needs to be positioned underneath the fiber array exactly the same way for each device. We also note that the actual location of the \emph{individual} fibers (more precisely, their cores) in the array varies slightly, leading to small differences in the measured transmission for different combinations of input and output fibers. Differences in the measured signal can also occur due to different gains of the photo detectors. Finally, the vertical distance between the sample and the fiber array can change when moving from one device to another. These issues are addressed by measuring different calibration devices in addition to the devices with the actual directional couplers. By carefully measuring their transmissions, the abovementioned factors can be quantified and corrected for to determine the insertion loss of the directional couplers.

For the measurements, every device (i.e., both the calibration and actual devices) is first placed underneath the fiber array and then its position is optimized using an automated procedure that allows for repeatability of the transmission (and thus of the position) at the percent level. This is done by maximizing the transmission between the input and the reference output to ensure that all four fibers are always at the same position above the grating couplers. Note that this would not necessarily be the case when maximizing the \emph{total} power $\Pref+P_t+P_c$ as the relative power at each output varies from device to device (as the splitting ratio is varied). After this optimization, the wavelength of the laser is swept and the transmission $T_{i\leftarrow2}(\lambda)$ at each of the three outputs [$i = 1$ (ref.), 3 (through), and 4 (cross); see Fig. \ref{fig:insertionloss}(a)] is measured.

The three different types of calibration devices are shown in Fig. \ref{fig:insertionloss}(a). The first (top) is a direct connection between the input (port 2) and the reference output (port 1). It thus allows calibration of $T_{1\leftarrow2}(\lambda)$. In the center calibration device the input light is split equally (using a Y-splitter) between the reference and port 3. The relative power at the reference port $P_1/\Pin$ is thus the product of the transmission through the Y-splitter $Y$ and $T_{1\leftarrow2}$. Since $T_{1\leftarrow2}(\lambda)$ is known from the first calibration device, this allows determination of $Y(\lambda)$. Likewise $P_3/\Pin$ is the product of $Y$ and $T_{3\leftarrow2}$, from which $T_{3\leftarrow2}(\lambda)$ can be determined. Finally, in the bottom device $\Pin$ is again split equally, but now between the reference and port 4, which yields the last transmission $T_{4\leftarrow2}(\lambda)$.\footnote{In practice, the transmissions are not determined in this step-by-step process, but instead all measured (logarithmic) transmission data is fitted simultaneously by using 15-th order polynomials in $\lambda$ for $T_{i\leftarrow2}(\lambda)$ and $Y(\lambda)$. This is done through linear fitting via LU decomposition. The final results do not depend significantly on the order of the polynomial as long as the order is high enough to follow the overall profile, but low enough to filter out the fine fringes. The small upturns near 1570 nm in the curves shown in Fig. \ref{fig:insertionloss}(a) are artifacts of the polynomial fits due the end of the data range.}
Figure \ref{fig:insertionloss}(b) shows the resulting transmission profiles $T_{i\leftarrow2}(\lambda)$ as well as the transmission through the Y-splitter. The latter is $Y \sim -0.8 \un{dB}$, largely independent of wavelength. Only small ($ < 0.5 \un{dB}$) differences in the transmission profiles $T_{i\leftarrow2}$ are visible\footnote{The maximum transmission in Fig. \ref{fig:insertionloss}(b) is $\sim -32 \un{dB}$. This value is lower than typical for our SiN devices ($\sim -15 \un{dB}$) since the fiber array was kept far away from the sample to prevent any accidental touching or scratching while scanning the entire chip. Note that none of the results presented here depend on the \emph{absolute} transmission.}. For example, the transmission through output 3 is about 1 dB larger than through port 4; the transmission through the reference port is in between the other two. If these differences were not accounted for, the apparent power coming out of the directional coupler, and thus the insertion loss, could be off by about half a dB.

After performing this calibration procedure the measured powers at each detector are corrected using the corresponding $T_{i\leftarrow2}(\lambda)$ before calculating the insertion loss $(P_c+P_t)/\Pref$.
Figure \ref{fig:insertionloss}(c) shows that the insertion loss of the 22 directional couplers has a mean value of 0.06 dB which is comparable to the standard deviation of 0.07 dB. The former indicates that $> 98\%$ of the photons that are incident on the beam splitter either go to the through or to the cross-over port; the insertion loss of the directional couplers is thus small.

\section{Beam splitter phase shift}
\label{sec:phase_shift}
\begin{figure}[tb]
\includegraphics{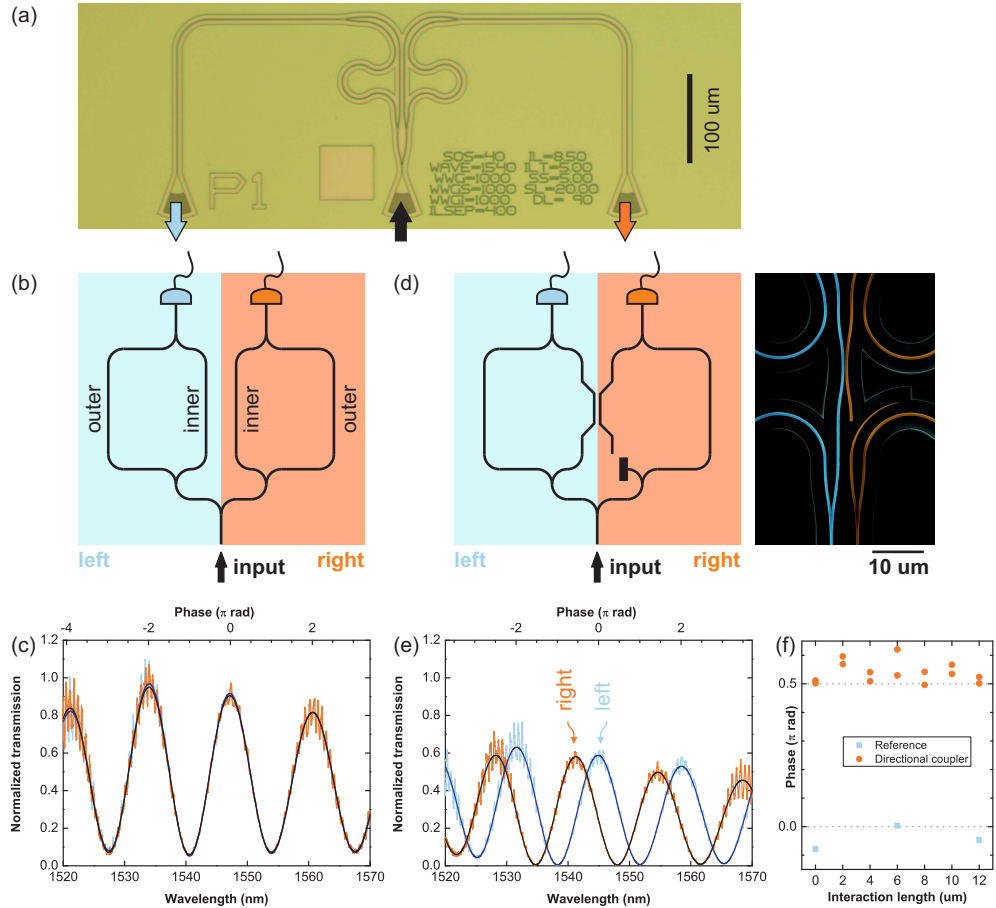}%
\caption{(a) optical micrograph of a device to determine the phase difference between the two outputs of a directional coupler
(b,d) schematics of the two type of interferometers. The left (right) MZI is indicated in blue (orange). The colorized electron micrograph on the right side of (d) shows the central area of a device with a directional coupler.
(c,e) MZI fringes measured at the two outputs of the device
(light blue, orange) together with fits (blue and black lines)
to determine the periods and wavelength shift. The phase of the fringes relative to the fringes of the left MZI is indicated on the top axis.
Panels (a) to (c) show a reference device, and (d) and (e)
are for a device with a directional coupler with an interaction length of $8 \un{\mu m}$.
(f) Phase difference between the two interferometers.}
\label{fig:directional_phase}
\end{figure}

Another important question about beam splitters and directional couplers is what relative phase difference it imprints on the output fields \cite{zeilinger_AJP_beamplitter}. In bulk optics this phase difference depends on the details of the beam splitter that is used. For example, upon reflection from a metallic beam-splitting surface a phase change of $\pi$ occurs with respect to the transmitted field. On the other hands, when the beam splitter is made using dielectric layers the situation is different. Moreover, the polarization of the incident light and coatings on the beam splitter can further complicate the relation between the output light fields \cite{mitchell_PRL_bell-state_filter_QPT}. For the actual design of the quantum circuitry it is essential to know this phase difference: compare for example the assignments of the ports of the CNOT gate in Ref. \cite{ralph_PRA_CNOT_coincidence_theory} with our design in Sec. \ref{sec:CNOT_ports}. These differences are solely due to the difference in the beam-splitter phase relations.

To find the optical phase difference between the outputs of our  integrated directional couplers we have made and measured the devices shown in Fig. \ref{fig:directional_phase}
which consist of two Mach-Zehnder interferometers (MZIs). Relative shifts between the fringes of these two unbalanced MZIs will be used to infer the phase difference between the outputs of the directional coupler. They work as follows: light from a central input is split equally and guided to the inputs of two MZIs (referred to as ``left'' and ``right''). At the input of each of these MZIs the light is split into two arms. After traveling through those arms (the shorter inner one and the longer outer arm; the latter acting as the reference arm of the MZI) the light is recombined at the output of each interferometer. Interference fringes, from which the phase difference can be determined, are measured by detecting the intensities at the outputs of the two unbalanced MZIs while sweeping the wavelength. These relative-phase measurements are performed on two different types of devices: one that consists of symmetric, uncoupled MZIs [Fig. \ref{fig:directional_phase}(b)], and a second type where the inner arms are connected via a directional coupler [Fig. \ref{fig:directional_phase}(d)]. As shown in Fig. \ref{fig:directional_phase}(c) the fringes of the symmetric, uncoupled device overlap, indicating that the phase difference between two interferometers, $\Delta \phi_L - \Delta \phi_R$, is small. These symmetric devices serve as validity checks of the method because they demonstrate that our interferometers have  identical phase-stable path lengths. In other words, the waveguides in different parts of our circuits can be controlled to have propagation-phase differences $\ll \pi$.

In the second type of device [Fig. \ref{fig:directional_phase}(d,e)] a directional coupler  connects the two inner arms of the interferometers. Light from the left inner arm is split between the left and right interferometers via the directional coupler; no light from the right side enters the directional coupler.

The phase of light that propagated through the outer arms is $\phi_{\mathrm{outer},L(R)}$. The phase of light in the inner arms $\phi_{\mathrm{inner},L(R)}$ consist of two parts: Firstly, the phase $\phi_{p,L(R)}$ accumulated by propagating from the bottom splitter to the directional coupler and from the directional coupler to the top of the interferometer where it combines with light that has traveled through the outer arms. The second contribution is the phase due to the directional coupler, which is given by the argument of the corresponding element of its scattering matrix. The phase differences are thus $\Delta \phi_{L(R)} = \phi_{p,L(R)} + \angle S'_{11}(\angle S'_{21}) - \phi_{\mathrm{outer},L(R)}$. Apart from the connections to the directional coupler all elements of the device are designed entirely symmetric and thus $\phi_{p,L} = \phi_{p,R}$ and $\phi_{\mathrm{outer},L} = \phi_{\mathrm{outer},R}$ as was verified with the symmetric devices. Now, a relative shift between the two fringes indicates that there is a phase difference between the two outputs of the directional coupler since $\Delta \phi_R - \Delta \phi_L = \angle S'_{21} - \angle S'_{11}$.

A shift between the fringes is indeed visible in Fig.
\ref{fig:directional_phase}(e); the fringes of the right
interferometer are at shorter wavelengths compared to
those of the left MZI. Since a full fringe period corresponds to a $2\pi$ change in $\Delta \phi_{L(R)}$, the wavelength shift can be converted to a relative phase shift\footnote{The convention where the optical field is
proportional to $\exp(-i\omega t)$ is used. In this case the
phase of light propagating in, say, the $+x$ direction is $\phi_p = +2\pi \neff x / \lambda$. Since the outer arms are the reference arms, the phase difference is defined as $\Delta \phi = \phi_{\mathrm{inner}} - \phi_{\mathrm{outer}}$. The outer arm is longer than the inner arm and hence $\Delta \phi < 0$. Since $\pderl{\phi_p}{\lambda} = -\phi_p/\lambda$ (without dispersion), the phase difference increases with increasing wavelength.} $\Delta \phi_R - \Delta \phi_L$ as indicated on the top axis of panels (c) and (e). By fitting the fringes, both the fringe period $\Delta \lambda$ and the wavelengths of the maxima $\lambda^{(L,R)}$ can be determined. Using the period and position of the fringes the phase difference is obtained: $\angle S'_{21} - \angle S'_{11} = 2\pi(\lambda_{L}^{\max} - \lambda_{R}^{\max})/\Delta \lambda$. The result for devices with different interaction lengths (i.e. splitting ratios) is shown in Fig. \ref{fig:directional_phase}(f). The symmetric devices are close to zero phase difference, whereas the directional-coupler MZIs are all close to $\pi/2$. The mean value of the latter is $+0.54\pi$ with a standard deviation of $0.04\pi$, indicating that the light picks up a factor of $i$ when crossing over to the other side of the beam splitter\footnote{Although in a general beam splitter any value of the phase difference is allowed (with constraints on their combination), for a symmetric design such as our directional coupler, only $-\pi/2$ and $+\pi/2$ are possible \cite{zeilinger_AJP_beamplitter}.}. This is also confirmed by the FEM simulation shown in the inset of Fig. \ref{fig:photonic}(a) where the field in the bottom waveguide is shifted by $\pi/2$ compared to the top one. Note that although Fig. \ref{fig:directional_phase}(f) shows that the phase difference is independent of the interaction length, this is only valid up to the first maximum in $C$ [cf. Fig. \ref{fig:photonic}(d)]. After each maximum, an additional $\pi$ phase shift is acquired and the phase difference would be $-\pi/2$. The phase shift would also change sign if the effective refractive index of the even mode were smaller than that of the odd mode [i.e. the reverse of the situation in Fig. \ref{fig:photonic}(c)]. In any case, the devices that are used in this work are thus accurately described by the scattering matrix
\begin{equation}
\mathbf{S'} = \left(\begin{array}{cc}
t & ic \\
ic & t
\end{array}\right),
\label{eq:beam_slitter}
\end{equation}
where $t$ is the transmission coefficient, and $c = (1-t^2)^{1/2}$ is the cross-over coefficient. By comparing Eq. (\ref{eq:beam_slitter}) with Eq. (\ref{eq:Smatrix}) we identify $t = S'_{11} = S'_{22}$ and $ic = S'_{12} = S'_{21} = iC^{1/2}$ when $S'_{11}$ is real and positive.

\section{Characterization of a quantum circuit}
\label{sec:CNOT_ports}
\begin{figure}[tb]
\includegraphics{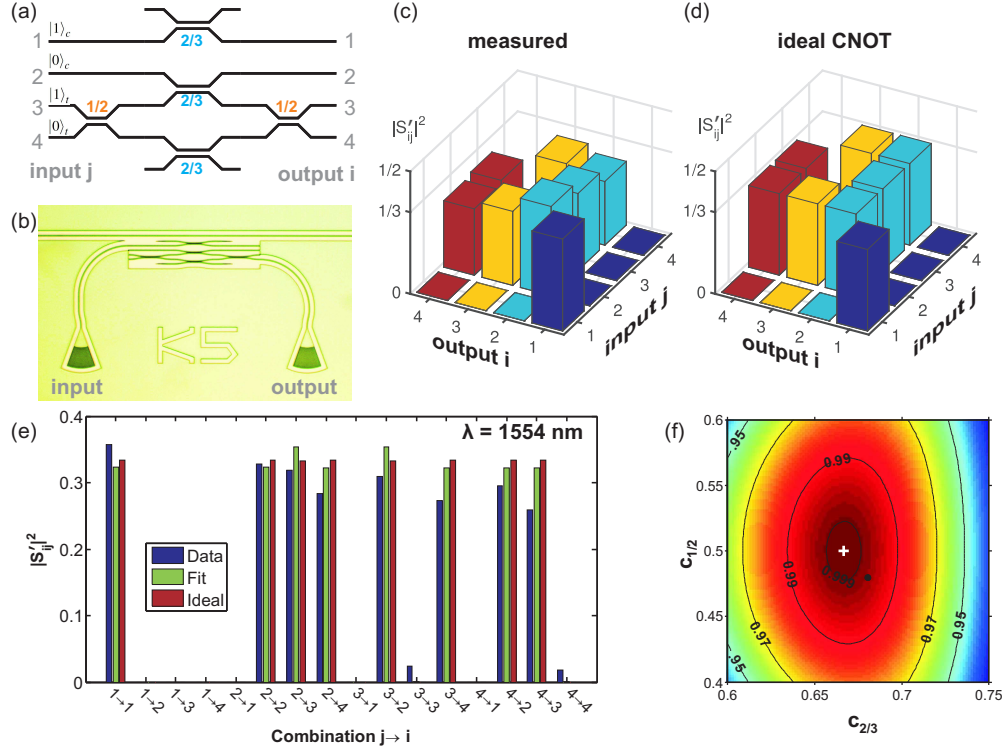}%
\caption{(a) Schematic of a CNOT gate for linear-optics quantum computation. The port number and the cross-over ratios of the directional couplers are indicated.
(b) Micrograph of a device to measure $|S'_{41}|^2$. The waveguide above the CNOT gate is used to locate the device as some combinations (including this one) have zero transmission. This auxiliary waveguide does not influence the actual CNOT circuit.
%
(c) Measured normalized transmission matrix and that for an ideal CNOT gate (d).
(e) Bar chart comparing the measured data [cf. (c)], the fit, and the transmission for an ideal CNOT gate [cf. (d)].
(f) Calculated fidelity, i.e. the probability of obtaining the right result after post selection vs. $C_{1/2}$ and $C_{2/3}$. The dot indicates the fitted values of (e), and the cross is located at the ideal values.}
\label{fig:CNOT_ports}
\end{figure}

The directional coupler characterization in Sections \ref{sec:directional} to \ref{sec:phase_shift} shows that with careful calibration the transmission through the device reveals the splitting ratio, phase shifts, and insertion loss. This approach also works for more complex photonic circuits \cite{rahimi-keshari_OE_linear_charactization}, such as, for example, the controlled-NOT circuit \cite{kok_RMP_LOQC} shown Fig. \ref{fig:CNOT_ports}. A CNOT gate is a quantum operation that acts on two qubits, the so-called control and target qubit. When CNOT gates are combined with single-qubit operations, any arbitrary unitary quantum operation on any number of qubits can be performed \cite{nielsen_book_quantum_computation}. In this section the transmission of classical light through a CNOT circuit is investigated. Using the obtained scattering matrix its quantum behavior is predicted.

A CNOT flips its target qubit $\ket{0}_t \leftrightarrow \ket{1}_t$ when the control qubit is in the logical quantum state $\ket{1}_c$, but leaves the target qubit unchanged when the control qubit is in the logical zero state $\ket{0}_c$. The linear-optics implementation of Fig. \ref{fig:CNOT_ports} relies on post selection, i.e., the outcome is nondeterministic, and the probability of success of the two-qubit operation is $1/9$ \cite{kok_RMP_LOQC}. Each qubit is encoded in the spatial-mode degree of freedom (i.e. waveguide) that a single photon takes. The four inputs are labeled $1 .. 4$ [see Fig. \ref{fig:CNOT_ports}(a)] and represent the logical states $\ket{1}_c$, $\ket{0}_c$, $\ket{1}_t$, and $\ket{0}_t$ respectively; the upper two ports thus form the control qubit whereas the lower two ports are the target qubit. In the case where the control qubit is in the $\ket{1}_c$ state (i.e. the photon is in waveguide 1), the target-qubit and control-qubit photon are in different parts of the circuit and thus do not interfere with each other. The target qubit is flipped (goes to the other waveguide) by the sequence of the two 50/50 beam splitters (unless it crosses over in a $C=2/3$ directional couplers; those cases are excluded via post selection). On the other hand, when the control qubit is in the $\ket{0}_c$ state, the two photons are in the same part of the circuit. Their interference at the central $C=2/3$ directional coupler ensures that after post selection \cite{kok_RMP_LOQC} the target photon is found in the same output waveguide as where it came from. In this case, the target qubit remains unchanged. The desired result is thus obtained for every value of the control qubit.

Since the operation of the CNOT relies on interference at the central beam splitter, it requires precise control over the splitting ratios of the directional couplers and phase stability of the optical paths. In an integrated circuit path length control is readily achieved as shown in Sec. \ref{sec:phase_shift}. Yet it is imperative to confirm that after fabrication of the device the circuits perform as designed, i.e. that the splitting ratios of all the directional couplers are correct. In a bulk optics experiment this can easily be verified by placing photodetectors in the paths of interest and measuring the relative powers. In a photonic circuit that is not as straightforward and one is often limited to using the input and output ports for characterization, although recently characterization of circuits using imaging of scattered light from the top has been reported \cite{metcalf_natcomm_multiphoton_circuit}.

As shown in Fig. \ref{fig:CNOT_ports}(a) our device has four inputs and four outputs and hence its (reduced) scattering matrix consists of 16 elements. Devices connecting different combinations of input $j$ and output ports $i$ are fabricated and their normalized transmission $|S'_{ij}|^2$ is measured. As an example, Fig. \ref{fig:CNOT_ports}(b) shows a device to measure $|S'_{41}|^2$ [which should be, and is, zero
as is readily verified from the schematic in Fig.
\ref{fig:CNOT_ports}(a)]. All combinations are shown in Fig.
\ref{fig:CNOT_ports}(c) and these should be compared with the
transmission matrix of an ideal CNOT shown in Fig. \ref{fig:CNOT_ports}(d). The latter has 8 elements which are 1/3 and the other 8 are zero. The experimental data looks similar, but some deviations that will be studied in more detail below can be seen.

By cascading the scattering matrices of all the individual components of the CNOT circuit, one can also calculate the
total transmission amplitudes as a function of the splitting ratios of the directional couplers that are ideally 1/2 and 2/3 [see Fig. \ref{fig:CNOT_ports}(a)]. The actual splitting ratios of the two types of directional couplers, $C_{1/2}$ and $C_{2/3}$ respectively, are determined by fitting the transmissions calculated via a model of the circuit to the experimental ones. Note that, given the reproducibility of the directional couplers demonstrated in Sec. \ref{sec:directional}, the coupling ratios of directional couplers with the same design are assumed to be identical in the fitting procedure. The model transmissions are calculated using the cascaded scattering matrices of the individual components of the circuit. The fitting is done for the devices of Fig. \ref{fig:CNOT_ports}, as well as for a second set of similar devices. The fits at $\lambda = 1554 \un{nm}$ [Fig. \ref{fig:CNOT_ports}(c)] give $C_{1/2} = 0.477~(0.480)$ and  $C_{2/3} = 0.676~(0.669)$ for the first (second) set. The small variation in the fit parameters between the two datasets again illustrates the excellent reproducibility of our integrated nanophotonic circuits. Also, the scattering matrices of the measured device and the fit are very similar: The mean of the absolute deviation between the measured and the fitted transmission is 0.02 for both datasets. This value corresponds to fluctuations of about 0.1 dB in the measured transmission and is comparable to the stochastic fluctuations measured in the transmission of calibration devices. The small deviations between the fit and measurement can thus be attributed to measurement uncertainties. The deviations between the fit and the transmission of an ideal CNOT are even smaller: 0.009 and 0.006 respectively. This shows the ability to build more complex quantum circuits.

As discussed above, the outcome of a CNOT gate as in Fig. \ref{fig:CNOT_ports}(a) is post selected when for both qubits a single photon is detected in just one of the two waveguides representing $\ket{0}$ and $\ket{1}$ \cite{kok_RMP_LOQC}. For the ideal design this occurs statistically in 1/9 of the attempts. After post selection, the ideal gate always yields the right outcome (its fidelity is 1). However, since the splitting ratios are in practice slightly different from the ideal values, the gate fidelities and the probabilities of getting a post-selected event will also be different \cite{ralph_PRA_CNOT_coincidence_theory}. The final state $\ket{\psi_{\mathrm{out}}}$ for an input state $\ket{\psi_{\mathrm{in}}}$ is given by $\ket{\psi_{\mathrm{out}}} = \hat U \ket{\psi_{\mathrm{in}}}$, where $\hat U$ is a unitary operator that is directly related to the classical scattering matrix of the linear circuit \cite{skaar_AmJP_quantum_linear_optics, leonhardt_JOB_linear_quantum_scattering_matrix} through the transformation of the creation operators of mode $i$: $\hat U a^\dagger_i \hat U^\dagger = \sum_j S_{ji} a^\dagger_j$. Using this formalism, the quantum behavior of the circuit can be predicted, in particular it can be used to estimate how often the circuit gives the right result (i.e. the outcome expected from the truth table) \cite{ralph_PRA_CNOT_coincidence_theory}. Figure \ref{fig:CNOT_ports}(f) shows a contour plot of the probability of the right outcome as a function of the two splitting ratios in the circuit, averaged over the input states $\ket{00} \equiv \ket{0}_c \otimes \ket{0}_t$, $\ket{01}$, $\ket{10}$, and $\ket{11}$. Exactly at the target values the fidelity is 1, but this value drops when moving away from this ideal case. For the fitted values $C_{1/2} = 0.477$ and $C_{2/3} = 0.676$ [as indicated by the dot in Fig. \ref{fig:CNOT_ports}(f)] the right outcome occurs 99.81\% of the time and the probability of obtaining a post-selectable output state is 10.95\% (compared to 1/9 $\approx$ 11.11\% for the ideal case). The nonideal splitting ratios thus only give a modest reduction of the fidelity of the CNOT gate. For the second dataset we obtain 99.92\%, and using the values from the directional couplers (Sec. \ref{sec:directional}) 99.993\%. We note that with such high values, in the future, the quantum process fidelity will probably be limited by the purity of the single photon source and not by the gate operation.


\section{Conclusion}
Targeting integrated linear optics quantum circuits in a silicon-nitride material platform, we have developed several essential components and established a detailed calibration procedure for each of them. By measuring the length dependence of the splitting ratio of directional couplers, the coupling length and offset length are determined, as well as their wavelength dependence. Also the phase difference between the transmitted and reflected light is studied, and the insertion loss is found to be small. By measuring and fitting the optical transmission between different pairs of ports of a CNOT gate, the splitting ratios of each of its constituents are found. The values obtained from our fitting procedure are close to the ideal values, indicating that the quantum operation should have a high fidelity. The fabrication process flow for these devices is entirely compatible with integration of SNSPDs and opto-electromechanical phase-shifters. This enables operation of monolithically integrated programmable quantum circuits with highly efficient superconducting detectors on the same chip. The next step is to build circuits that combine initialization, computation, tomography, and detection, and use nonclassical light to perform the desired quantum operations. With the integrated approach, more complex operations are certainly possible and the ultimate goal of a universal quantum computer based on LOQC may be within reach.

\section*{Acknowledgments} This work was
partly funded by the Packard Foundation. C.S. acknowledges financial support from the Deutsche Forschungsgemeinschaft (SCHU 2871/2-1). H.X.T. acknowledges support from a career award from National Science Foundation. Facilities used were supported by Yale Institute for Nanoscience and Quantum Engineering and NSF MRSEC DMR 1119826.
\end{document}